\documentclass[onecolumn]{aastex63}
\usepackage[utf8]{inputenc}
\usepackage{epsfig}
\usepackage{scalerel}

\shortauthors{J. Li et al.}
\shorttitle{Errata of Li's paper}

\begin{document}

\title{Errata on the Calculation of Hot Gas Properties in a Few Li Jiang-Tao's Papers}

\author[0000-0001-6239-3821]{Jiang-Tao Li}
\affiliation{Department of Astronomy, University of Michigan, 311 West Hall, 1085 S. University Ave, Ann Arbor, MI, 48109-1107, U.S.A.}

\author{Rui Huang}
\affiliation{Department of Astronomy, Tsinghua University, Beijing 100084, China}
\affiliation{Department of Astronomy, University of Michigan, 311 West Hall, 1085 S. University Ave, Ann Arbor, MI, 48109-1107, U.S.A.}

\correspondingauthor{Jiang-Tao Li}
\email{pandataotao@gmail.com}

\begin{abstract}
This is a combination of the errata of seven papers published between 2008 and 2016 with Jiang-Tao Li (JTL) as the first author. All the problems are caused by two mistakes in the original scripts written by JTL used to calculate the physical parameters of the hot gas from X-ray spectral analysis with a thermal plasma code. The mistakes will result in an overestimate of some parameters, such as the electron number density and hot gas mass by a factor of $\sqrt{10}\approx3.162$, and an overestimate of the thermal pressure by a factor of  $\approx2.725$. JTL apologizes to the community for the inconvenience caused by these mistakes. We present an update on the text, numbers, figures, and tables of all the seven papers affected by these mistakes. Other papers led by JTL or co-authored papers are not affected.
\end{abstract}



\section{The Mistakes} \label{sec:Mistakes}

Jiang-Tao Li (JTL) made a mistake in unit conversion when calculating the emission measure and electron number density from the normalization of a fitted thermal plasma model (e.g., MEKAL or APEC). Below we describe where the mistakes come from.

The normalization of a thermal plasma model (e.g., MEKAL or APEC) is often defined as:
\begin{equation}\label{equ:norm}
norm=\frac{10^{-14}}{4\pi [D_{A}(1+z)]^2}\int n_{e}n_{H}dV,
\end{equation}
where $D_{A}$ is the angular diameter distance to the source in unit of cm, $n_{e}$ and $n_{H}$ are the number densities of electron and Hydrogen in unit of $\rm cm^{-3}$.

We convert the measured $norm$ to the volume emission measure $EM$ and then to the Hydrogen number density and other physical parameters. The original formulae in the script are:
\begin{equation}\label{equ:EM1}
EM=4\pi~norm~d^2/pc,
\end{equation}
\begin{equation}\label{equ:nH}
n_{H}=\sqrt{EM/V_{emit}/\mathcal{R}_{n_e n_H}},
\end{equation}
where $d$ is the distance to a local galaxy in Mpc (we do not consider the $1+z$ term for local galaxies), $pc=3.086$ is in unit of $10^{18}\rm~cm$, $V_{emit}$ is the volume of the X-ray emitting region in unit of $\rm kpc^3$, and $\mathcal{R}_{n_e n_H}$ is the electron to Hydrogen number ratio depending on the metallicity and ionization state of the gas. In some early papers, we have assumed $\mathcal{R}_{n_e n_H}=1$ \citep{Li08,Li09,Li13a}, while in later papers, $\mathcal{R}_{n_e n_H}$ has been calculated based on the measured metallicity \citep{Li15a,Li16a,Li16b}. As $\mathcal{R}_{n_e n_H}$ is very close to 1 ($\sim1.2$ for fully ionized solar abundance gas) and the actual value of it does not affect the result significantly, we still follow the original assumption of it in the corresponding papers. Here the derived Hydrogen number density is in unit of $f^{-1/2}\rm~cm^{-3}$, where $f$ is the unknown volume filling factor of the hot gas. The exact forms of the above equations depend on the adopted unit of the parameters. In Eq.~\ref{equ:EM1}, JTL has missed a factor of 10. The correct formula should actually be:
\begin{equation}\label{equ:EM2}
EM=4\pi~norm~d^2/pc/10.
\end{equation} 

This mistake will cause an overestimate of the emission measure by a factor of 10 and an overestimate of the Hydrogen and electron number densities by a factor of $\sqrt{10}$. All the related parameters, such as the thermal pressure, the mass and thermal energy of hot gas, are also overestimated by a factor of $\sqrt{10}$. 

Furthermore, JTL also adopted an inaccurate formula when calculating the thermal pressure of hot gas (in unit of $f^{-1/2}\rm~eV/cm^3$):
\begin{equation}\label{equ:Pthem1}
P_{them}=10^7n_e k_BT, 
\end{equation} 
where $k_B$ is the Boltzmann constant ($8.6173325\times10^{-5}\rm~eV/K$), $n_e$ is in unit of $f^{-1/2}\rm~cm^{-3}$, and $T$ is the hot gas temperature in unit of keV. The formula itself is not wrong, but contains an inaccurate relation converting the temperature unit from eV to Kelvin (JTL has simply assumed $1\rm~eV=10^4~K$). A more accurate form of Eq.~\ref{equ:Pthem1} should be:
\begin{equation}\label{equ:Pthem2}
P_{them}=10^3n_e k_BT, 
\end{equation} 
where $k_BT$ as a single parameter, instead of $T$, is in unit of keV. This mistake will result in an underestimate of the thermal pressure by a factor of $10^4/k_B\approx1.160$. 

The combined effect of the above two mistakes is an overestimate of the electron and Hydrogen number densities, the mass and thermal energy of the hot gas by a factor of $\sqrt{10}\approx3.162$, and an overestimate of the thermal pressure by a factor of  $\sqrt{10}/1.160\approx2.725$. The case of the radiative cooling timescale $t_{cool}$ is more complicated as it was estimated in different ways in different papers. In some early papers \citep{Li08,Li09,Li13a}, $t_{cool}$ is obtained by directly dividing the thermal energy with the observed X-ray luminosity, so it is $\propto n_e$ and also overestimated by a factor of $\approx3.162$. In \citet{Li14}, we also present the radiative cooling rate $\dot{M}_{cool}$ in Fig.1b of that paper, which is obtained by dividing the total hot gas mass in the CGM $M_{hot}$ with $t_{cool}$. As $M_{hot}$ and $t_{cool}$ in this paper are both overestimated by a factor of $\sqrt{10}$, $\dot{M}_{cool}$ in Fig.1b of \citet{Li14} and the related discussions are not affected by this mistake. We therefore do not issue an erratum of \citet{Li14}. However, in some later papers such as \citet{Li16b}, $t_{cool}$ is obtained from the cooling function, so it is $\propto n_e^{-1}$ and has been underestimated by a factor of $\approx3.162$. We also calculated the radiative cooling rate $\dot{M}_{cool}$ in these later papers, which is $\propto n_e^2$ so has been overestimated by a factor of 10. Other directly measured hot gas parameters, such as the X-ray luminosity, and sometimes temperature and metallicity, are not affected by these mistakes. We also want to emphasize that in the two latter CGM-MASS papers \citep{Li17,Li18}, we use a different method to calculate the derived physical parameters of the hot gas from those adopted in CGM-MASS~I \citep{Li16b}. This new method directly convert the observed X-ray counts rate, metallicity, and radial intensity profile to the radial distribution of physical parameters, so is not affected by the above mistakes. There is thus no erratum of \citet{Li17,Li18}.

These mistakes were not noticed by JTL, as the deviations of the estimated hot gas parameters are not quite significant, and the estimate itself is often highly uncertain due to the large errors of the measurement and many assumptions (e.g., the unknown volume filling factor $f$). A graduate student, Rui Huang (RH), working with JTL has recently found the mistakes in a careful examination of the early scripts written by JTL. As the scripts have been used for many years without an update, seven of JTL's first author papers published from 2008 to 2016 have been more or less affected by these mistakes. JTL would like to apologize to the community for the inconvenience caused by these mistakes. We herein present errata of all these seven papers, with the corrected text marked in {\color{red}red}. All these errata will also be separately submitted to the corresponding journals. 

\section{Erratum: ``C\lowercase{handra observation of the edge-on spiral} NGC 5775: \lowercase{probing the hot galactic disc/halo connection}'' (\href{https://doi.org/10.1111/j.1365-2966.2008.13749.x}{2008, MNRAS, 390, 59})} \label{sec:Li08}

$\bullet$ The $EM$ in Table~3 of \citet{Li08} should be updated to those in Table~\ref{Li08:specfitpara} of this erratum.

$\bullet$ Physical parameters in Table~4 of \citet{Li08} should be updated to those in Table~\ref{Li08:halopara} of this erratum.

$\bullet$ In section~4.3, 5th paragraph of \citet{Li08}. The following sentence should be changed to:

``The hot-gas pressure in the halo is {\color{red}$\sim3.8\times10^{-13}\rm~dyn~cm^{-2}$ for each of the two components}, derived from the hot-gas parameters listed in Table~4.''

\begin{deluxetable}{lrrrrrrrr}[!h]
  \tablecaption{Spectral Fitting Parameters (Table~3 of \citet{Li08})}
  \tabletypesize{\footnotesize}
  \tablehead{
  \colhead{Region} & Model & $N_H$ & $T_{Low}$ & $EM_{Low}$ & $T_{High}$ & $EM_{High}$ & $\Gamma$ & $\chi^2/d.o.f.$ \\
  \colhead{} & & $(10^{20}\rm~cm^{-2})$ & (keV) & ($\rm cm^{-6}kpc^3$) & (keV) & ($\rm cm^{-6}kpc^3$) & & \\
  \colhead{(1)} & (2) & (3) & (4) & (5) & (6) & (7) & (8) & (9)
    }
\startdata
ULX & PL & $286_{-37}^{+50}$ & - & - & - & - & $1.82_{-0.22}^{+0.31}$ & 50.6/44 \\
Disk & Mekal+PL & $53.9_{-3.9}^{+10.5}$ & $0.19_{-0.01}^{+0.01}$ & {\color{red}0.131} & - & - & 1.34 (fixed) & 92.9/74 \\
Halo & Mekal+Mekal & 3.48 (fixed) & 0.17 ($<0.19$) & {\color{red}0.0047} & $0.57_{-0.015}^{+0.015}$ & {\color{red}0.0017} & -& 70.4/69
\enddata
\tablecomments{(1) Regions from which spectra were extracted; (2) Mekal = 
XSPEC MEKAL thermal plasma model, PL = Power Law; (3) Hydrogen
absorption column density; (4)(6) Temperatures of the thermal
components; (5)(7) Emission measure of the thermal components; (8)
Photon index of the PL component; (9) Statistics of the spectral
fit. Note that the last column is calculated with the sky
background, while the spectra shown in Fig. 8
is after
sky background subtraction.}\label{Li08:specfitpara}
  \end{deluxetable}
  \vfill

\begin{deluxetable}{lrrrrrrrr}[!h]
  \tabletypesize{\footnotesize}
  \tablecaption{Hot Halo Gas Parameters (Table~4 of \citet{Li08})}
  \tablewidth{0pt}
  \tablehead{
  \colhead{Parameters} & Low T Component & High T Component \\
    }
\startdata
Volume Filling Factor & 0.2 & 0.8 \\
Number Density ($10^{-3}~\rm cm^{-3}$) & {\color{red}1.4} & {\color{red}0.41} \\
Mass ($10^8~\rm M_\odot$) & {\color{red}0.76} & {\color{red}0.92} \\
Thermal Energy ($10^{56}\rm~ergs$) & {\color{red}0.44} & {\color{red}1.7} \\
Cooling Time Scale ($10^9\rm~yr$) & {\color{red}0.63} & {\color{red}4.1} \\
\enddata

\tablecomments{These quantities are derived from the parameters
listed in Table~3, 
assuming that the filling factor
of the hot gas is $f_{high}$ ($f_{low}$) for the high (low)
temperature component with $f_{high}+f_{low}\sim1$, and that the hot
gas is located in a cylinder with a diameter of
$D_{25}$.}\label{Li08:halopara}
  \end{deluxetable}
  \vfill

\section{Erratum: ``D\lowercase{ynamic} S0 \lowercase{Galaxies: a Case Study of} NGC 5866'' (\href{https://iopscience.iop.org/article/10.1088/0004-637X/706/1/693}{2009, A\lowercase{p}J, 706, 693})} \label{sec:Li09}

$\bullet$ In section 3.2, the last paragraph of \citet{Li09}, the following sentence should be changed to:

``The corresponding emission measure of the hot gas is {\color{red}$\sim0.007\rm~cm^{-6}~kpc^3$}.''

$\bullet$ In section 4.2, the first paragraph of \citet{Li09}, the following sentence should be changed to:

``we estimate the mean {\color{red}electron} density, thermal pressure, and radiative cooling timescale of the hot gas as {\color{red}$\sim0.021\rm~cm^{-3}$, $\sim3.6\times10^{-12}\rm~dyne~cm^{-2}$, and $\sim4.7\times10^7\rm~yr$}. The total hot gas mass is  {\color{red}$\sim10^7\rm~M_\odot$}.''

$\bullet$ In section 4.3, the first paragraph of \citet{Li09}, the estimate of the warm gas properties is based on the assumption of pressure balance between hot gas and warm gas, so also indirectly affected by the mistakes. However, as the pressure balance assumption itself is highly uncertain, and the estimate of the warm gas properties is just accurate on order of magnitude, we will not change the numbers in this paragraph. The following discussions will not be affected.



\section{Erratum: ``C\lowercase{handra survey of nearby highly inclined disc galaxies} - I. X-\lowercase{ray measurements of galactic coronae}" (\href{https://doi.org/10.1093/mnras/sts183}{2013, MNRAS, 428, 2085})} \label{sec:Li13a}

The following tables should be updated accordingly. We also slightly changed the format of the table, i.e., use $value \pm limit$ to replace $value_{-lower}^{+upper}$ when the lower and upper limits of a parameter equal to each other.

$\bullet$ Physical parameters in Table~7 of \citet{Li13a} should be updated to those in Table~\ref{L13atable:1Tspec} of this erratum.

$\bullet$ Physical parameters in Table~8 of \citet{Li13a} should be updated to those in Table~\ref{Li13atable:specpara1Tmetal} of this erratum.

$\bullet$ Physical parameters in Table~9 of \citet{Li13a} should be updated to those in Table~\ref{Li13atable:specpara2T} of this erratum.

\begin{figure}[!h]
\begin{center}
\epsfig{figure=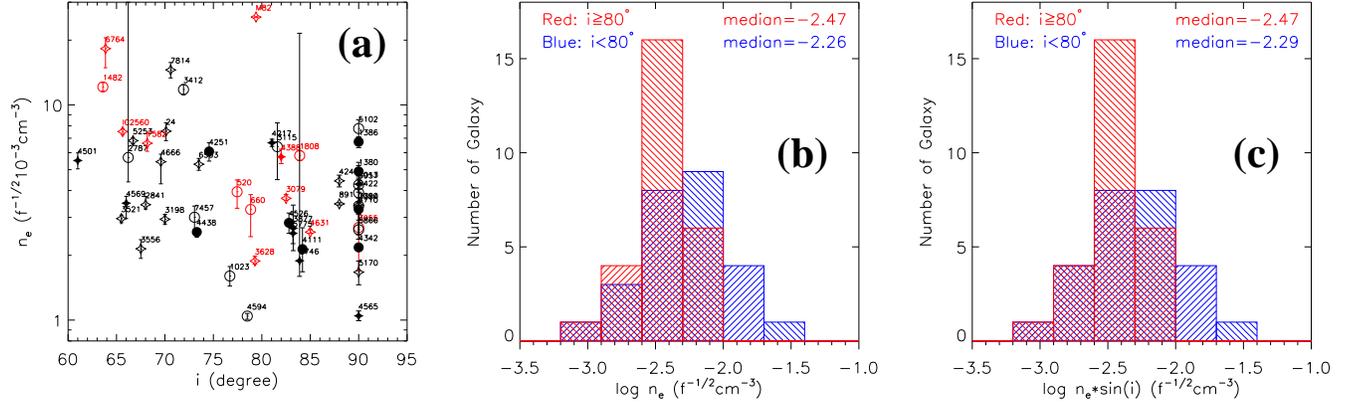,width=1.0\textwidth,angle=0, clip=}
\caption{(Fig.10 of \citet{Li13a}) (a) Hot gas density ($n_e$) vs. inclination angle ($i$), where the symbols have the same meanings as those in Fig.~8.
(b) $n_e$ distributions of highly ($i\geq80^\circ$) and moderately ($i<80^\circ$) inclined galaxies, where the median value of $\log n_e$ for each subsamples is given at the upper right corner. (c) Similar to (b), but with a simple geometry correction for the volume used in the calculation of $n_e$.}\label{Li13afig:test_incli}
\end{center}
\end{figure}

$\bullet$ Fig.~10 of \citet{Li13a} should be updated to Fig.~\ref{Li13afig:test_incli} of this erratum. An updated calculation indicates the correlation coefficient of the $i-n_e$ relation is $|r_s|<0.3$, so we remove it from the plot. Furthermore, in \S5.2.2, the following sentence  

``As an example (the most significant dependence on $i$), Fig.~10(a) shows that $n_e$ seems to be slightly correlated with $i$; but the correlation is only marginal ($-0.31\pm0.12$) and probably dominated by only a few galaxies with high $n_e$ values.''

should be changed to:

``{\color{red}An example is shown in Fig.~10(a), where $n_e$ does not show a significant correlation with $i$.}''

\begin{deluxetable}{lccccccc}[!h]
\centering
\tiny 
  \tabletypesize{\tiny}
  \tablecaption{Hot Gas Properties from the 1-T Model Fits (Table~7 of \citet{Li13a})}
  \tablewidth{0pt}
  \tablehead{
 \colhead{Name} & \colhead{$L_{hot}$} & \colhead{$T_{hot}$} & \colhead{$EM$} & \colhead{$n_e$} & \colhead{$M_{hot}$} & \colhead{$E_{hot}$} & \colhead{$t_{cool}$} \\
    & ($10^{38}\rm ergs/s$) & (keV) & ($10^{-2}\rm cm^{-6}kpc^3$) & ($f^{-1/2}10^{-3}\rm cm^{-3}$) & ($f^{1/2}10^8\rm M_\odot$) & ($f^{1/2}10^{55}\rm ergs$) & ($f^{1/2}\rm Gyr$) \\
   & (1) & (2) & (3) & (4) & (5) & (6) & (7)
}
\startdata
IC2560 & {\color{red}$66.34\pm 4.01$} & 0.3 & {\color{red}$2.35\pm 0.14$} & {\color{red}$2.52\pm 0.08$} & {\color{red}$2.31\pm 0.07$} & {\color{red}$13.2\pm 4.4$} & {\color{red}$0.63\pm 0.21$} \\
M82 & $68.76_{-0.30}^{+0.29}$ & {\color{red}$0.611\pm 0.003$} & {\color{red}$1.79\pm 0.01$} & {\color{red}$9.48\pm 0.02$} & {\color{red}$0.4669\pm 0.0010$} & {\color{red}$5.43\pm 0.03$} & {\color{red}$0.251\pm 0.002$} \\
NGC0024 & {\color{red}$1.70\pm 0.33$} & 0.3 & {\color{red}$0.049\pm 0.009$} & {\color{red}$4.25\pm 0.41$} & {\color{red}$0.028\pm 0.003$} & {\color{red}$0.16\pm 0.06$} & {\color{red}$0.30\pm 0.12$} \\
NGC0520 & $19.06_{-6.22}^{+3.71}$ & $0.29_{-0.03}^{+0.05}$ & {\color{red}$0.36_{-0.11}^{+0.10}$} & {\color{red}$1.76_{-0.28}^{+0.24}$} & {\color{red}$0.50_{-0.08}^{+0.07}$} & {\color{red}$2.8_{-0.5}^{+0.6}$} & {\color{red}$0.46_{-0.17}^{+0.13}$} \\
NGC0660 & $11.73_{-2.03}^{+3.05}$ & $0.52_{-0.13}^{+0.11}$ & {\color{red}$0.15_{-0.08}^{+0.05}$} & {\color{red}$1.46_{-0.37}^{+0.25}$} & {\color{red}$0.25_{-0.06}^{+0.04}$} & {\color{red}$2.5_{-0.9}^{+0.7}$} & {\color{red}$0.68_{-0.27}^{+0.26}$} \\
NGC0891 & $22.26_{-0.51}^{+0.50}$ & {\color{red}$0.34\pm 0.01$} & {\color{red}$0.57\pm 0.01$} & {\color{red}$1.26\pm 0.02$} & {\color{red}$1.11\pm 0.01$} & {\color{red}$7.1\pm 0.1$} & {\color{red}$1.01\pm 0.03$} \\
NGC1023 & $2.81_{-0.70}^{+0.59}$ & $0.26_{-0.02}^{+0.03}$ & {\color{red}$0.055_{-0.011}^{+0.012}$} & {\color{red}$0.71_{-0.07}^{+0.08}$} & {\color{red}$0.19\pm 0.02$} & {\color{red}$0.94_{-0.12}^{+0.14}$} & {\color{red}$1.07_{-0.30}^{+0.27}$} \\
NGC1380 & $38.95_{-3.64}^{+3.35}$ & {\color{red}$0.33\pm 0.02$} & {\color{red}$0.65_{-0.06}^{+0.09}$} & {\color{red}$2.19_{-0.11}^{+0.14}$} & {\color{red}$0.73_{-0.04}^{+0.05}$} & {\color{red}$4.5_{-0.3}^{+0.4}$} & {\color{red}$0.37_{-0.04}^{+0.05}$} \\
NGC1386 & $14.86_{-1.97}^{+1.17}$ & {\color{red}$0.26\pm 0.01$} & {\color{red}$0.29\pm 0.03$} & {\color{red}$3.00_{-0.18}^{+0.16}$} & {\color{red}$0.24\pm 0.01$} & {\color{red}$1.18\pm 0.09$} & {\color{red}$0.25_{-0.04}^{+0.03}$} \\
NGC1482 & $37.21_{-3.31}^{+3.05}$ & {\color{red}$0.38\pm 0.03$} & {\color{red}$0.57\pm 0.05$} & {\color{red}$4.09_{-0.18}^{+0.19}$} & {\color{red}$0.35\pm 0.02$} & {\color{red}$2.5\pm 0.2$} & {\color{red}$0.21\pm 0.03$} \\
NGC1808 & $9.09_{-1.02}^{+0.92}$ & $0.58_{-0.06}^{+0.03}$ & {\color{red}$0.12\pm 0.01$} & {\color{red}$1.77\pm 0.09$} & {\color{red}$0.168\pm 0.009$} & {\color{red}$1.8_{-0.2}^{+0.1}$} & {\color{red}$0.64_{-0.10}^{+0.08}$} \\
NGC2787 & $1.77_{-1.32}^{+1.85}$ & $0.18_{-0.18}^{+0.11}$ & {\color{red}$0.12_{-0.06}^{+3.21}$} & {\color{red}$2.54_{-0.58}^{+33.46}$} & {\color{red}$0.12_{-0.03}^{+1.56}$} & {\color{red}$0.41_{-0.43}^{+5.47}$} & {\color{red}$0.74_{-0.94}^{+9.84}$} \\
NGC2841 & $19.37_{-3.17}^{+2.77}$ & $0.41_{-0.04}^{+0.07}$ & {\color{red}$0.47_{-0.04}^{+0.09}$} & {\color{red}$1.53_{-0.07}^{+0.14}$} & {\color{red}$0.76_{-0.03}^{+0.07}$} & {\color{red}$6.0_{-0.6}^{+1.2}$} & {\color{red}$0.98_{-0.19}^{+0.24}$} \\
NGC3079 & $65.90_{-3.77}^{+3.61}$ & {\color{red}$0.51\pm 0.02$} & {\color{red}$1.64_{-0.07}^{+0.14}$} & {\color{red}$1.52_{-0.03}^{+0.06}$} & {\color{red}$2.66_{-0.05}^{+0.11}$} & {\color{red}$25.8_{-1.2}^{+1.6}$} & {\color{red}$1.25_{-0.09}^{+0.10}$} \\
NGC3115 & $0.41_{-0.19}^{+0.13}$ & $0.08_{-0.08}^{+0.04}$ & {\color{red}$0.23_{-0.14}^{+0.13}$} & {\color{red}$2.85_{-0.85}^{+0.83}$} & {\color{red}$0.20\pm 0.06$} & {\color{red}$0.31_{-0.32}^{+0.17}$} & {\color{red}$2.36_{-2.69}^{+1.49}$} \\
NGC3198 & {\color{red}$15.62\pm 1.69$} & 0.3 & {\color{red}$0.42\pm 0.05$} & {\color{red}$1.31\pm 0.07$} & {\color{red}$0.79\pm 0.04$} & {\color{red}$4.5\pm 1.5$} & {\color{red}$0.91\pm 0.32$} \\
NGC3384 & {\color{red}$7.18\pm 1.26$} & 0.3 & {\color{red}$0.15\pm 0.03$} & {\color{red}$1.31\pm 0.12$} & {\color{red}$0.28\pm 0.02$} & {\color{red}$1.6\pm 0.5$} & {\color{red}$0.69\pm 0.27$} \\
NGC3412 & {\color{red}$9.82\pm 1.07$} & 0.3 & {\color{red}$0.26\pm 0.03$} & {\color{red}$5.25\pm 0.28$} & {\color{red}$0.124\pm 0.007$} & {\color{red}$0.71\pm 0.24$} & {\color{red}$0.23\pm 0.08$} \\
NGC3521 & $20.24_{-1.37}^{+1.35}$ & $0.36_{-0.02}^{+0.03}$ & {\color{red}$0.52\pm 0.04$} & {\color{red}$1.32_{-0.06}^{+0.05}$} & {\color{red}$0.97\pm 0.04$} & {\color{red}$6.6_{-0.4}^{+0.6}$} & {\color{red}$1.04_{-0.10}^{+0.12}$} \\
NGC3556 & $5.73_{-1.16}^{+1.01}$ & {\color{red}$0.33\pm 0.02$} & {\color{red}$0.15\pm 0.03$} & {\color{red}$0.80\pm 0.08$} & {\color{red}$0.45_{-0.04}^{+0.05}$} & {\color{red}$2.8\pm 0.3$} & {\color{red}$1.56_{-0.36}^{+0.33}$} \\
NGC3628 & $24.45_{-2.06}^{+1.86}$ & {\color{red}$0.32\pm 0.01$} & {\color{red}$0.61_{-0.04}^{+0.06}$} & {\color{red}$0.75_{-0.02}^{+0.04}$} & {\color{red}$2.01_{-0.07}^{+0.11}$} & {\color{red}$12.4_{-0.7}^{+0.8}$} & {\color{red}$1.61\pm 0.16$} \\
NGC3877 & $1.21_{-0.68}^{+0.61}$ & $0.30_{-0.06}^{+0.05}$ & {\color{red}$0.030_{-0.013}^{+0.017}$} & {\color{red}$0.78_{-0.17}^{+0.22}$} & {\color{red}$0.097_{-0.021}^{+0.027}$} & {\color{red}$0.56_{-0.16}^{+0.19}$} & {\color{red}$1.47_{-0.93}^{+0.88}$} \\
NGC3955 & $10.77_{-6.27}^{+3.67}$ & $0.31_{-0.05}^{+0.29}$ & {\color{red}$0.18_{-0.14}^{+0.11}$} & {\color{red}$1.20_{-0.44}^{+0.36}$} & {\color{red}$0.38_{-0.14}^{+0.11}$} & {\color{red}$2.2_{-0.9}^{+2.2}$} & {\color{red}$0.65_{-0.46}^{+0.68}$} \\
NGC3957 & {\color{red}$19.68\pm 2.95$} & 0.3 & {\color{red}$0.42\pm 0.06$} & {\color{red}$1.89\pm 0.14$} & {\color{red}$0.55\pm 0.04$} & {\color{red}$3.1\pm 1.1$} & {\color{red}$0.50\pm 0.19$} \\
NGC4013 & {\color{red}$14.47\pm 1.39$} & 0.3 & {\color{red}$0.39\pm 0.04$} & {\color{red}$1.59\pm 0.08$} & {\color{red}$0.61\pm 0.03$} & {\color{red}$3.5\pm 1.2$} & {\color{red}$0.77\pm 0.27$} \\
NGC4111 & $3.74_{-2.13}^{+1.36}$ & $0.44_{-0.10}^{+0.12}$ & {\color{red}$0.052_{-0.023}^{+0.027}$} & {\color{red}$0.76_{-0.16}^{+0.20}$} & {\color{red}$0.17\pm 0.04$} & {\color{red}$1.4_{-0.5}^{+0.6}$} & {\color{red}$1.23_{-0.80}^{+0.65}$} \\
NGC4217 & {\color{red}$20.35\pm 1.59$} & 0.3 & {\color{red}$0.55\pm 0.04$} & {\color{red}$2.08\pm 0.08$} & {\color{red}$0.65\pm 0.03$} & {\color{red}$3.7\pm 1.3$} & {\color{red}$0.58\pm 0.20$} \\
NGC4244 & {\color{red}$1.08\pm 0.14$} & 0.3 & {\color{red}$0.030\pm 0.004$} & {\color{red}$1.98\pm 0.12$} & {\color{red}$0.037\pm 0.002$} & {\color{red}$0.21\pm 0.07$} & {\color{red}$0.63\pm 0.23$} \\
NGC4251 & {\color{red}$23.05\pm 4.52$} & 0.3 & {\color{red}$0.45\pm 0.09$} & {\color{red}$2.71\pm 0.27$} & {\color{red}$0.41\pm 0.04$} & {\color{red}$2.3\pm 0.8$} & {\color{red}$0.32\pm 0.13$} \\
NGC4342 & $85.89_{-2.92}^{+2.89}$ & {\color{red}$0.53\pm 0.02$} & {\color{red}$1.15\pm 0.04$} & {\color{red}$1.23\pm 0.02$} & {\color{red}$2.32\pm 0.04$} & {\color{red}$23.4\pm 0.8$} & {\color{red}$0.87\pm 0.04$} \\
NGC4388 & $44.31_{-5.55}^{+7.00}$ & $0.61_{-0.05}^{+0.04}$ & {\color{red}$1.17_{-0.16}^{+0.20}$} & {\color{red}$2.11_{-0.15}^{+0.18}$} & {\color{red}$1.37_{-0.10}^{+0.12}$} & {\color{red}$15.8\pm 1.8$} & {\color{red}$1.13_{-0.19}^{+0.22}$} \\
NGC4438 & $52.76_{-3.95}^{+3.65}$ & {\color{red}$0.52\pm 0.03$} & {\color{red}$0.72_{-0.08}^{+0.05}$} & {\color{red}$1.06_{-0.06}^{+0.04}$} & {\color{red}$1.69_{-0.09}^{+0.06}$} & {\color{red}$16.9_{-1.4}^{+1.3}$} & {\color{red}$1.01_{-0.11}^{+0.10}$} \\
NGC4501 & $32.12_{-6.13}^{+5.89}$ & $0.56_{-0.07}^{+0.05}$ & {\color{red}$0.82\pm 0.14$} & {\color{red}$1.69\pm 0.14$} & {\color{red}$1.20\pm 0.10$} & {\color{red}$12.8_{-1.9}^{+1.7}$} & {\color{red}$1.26_{-0.31}^{+0.28}$} \\
NGC4526 & $8.84_{-2.05}^{+1.83}$ & $0.27_{-0.02}^{+0.04}$ & {\color{red}$0.17\pm 0.04$} & {\color{red}$0.96_{-0.10}^{+0.11}$} & {\color{red}$0.44\pm 0.05$} & {\color{red}$2.2_{-0.3}^{+0.4}$} & {\color{red}$0.80_{-0.21}^{+0.22}$} \\
NGC4565 & $8.87_{-0.84}^{+0.87}$ & $0.36_{-0.02}^{+0.04}$ & {\color{red}$0.23\pm 0.02$} & {\color{red}$0.47_{-0.02}^{+0.03}$} & {\color{red}$1.21_{-0.06}^{+0.07}$} & {\color{red}$8.3_{-0.7}^{+0.9}$} & {\color{red}$2.97_{-0.37}^{+0.44}$} \\
NGC4569 & $11.43_{-4.45}^{+1.97}$ & {\color{red}$0.56\pm 0.04$} & {\color{red}$0.30_{-0.09}^{+0.04}$} & {\color{red}$1.33_{-0.20}^{+0.10}$} & {\color{red}$0.56_{-0.08}^{+0.04}$} & {\color{red}$6.0_{-1.0}^{+0.6}$} & {\color{red}$1.65_{-0.70}^{+0.34}$} \\
NGC4594 & $20.66_{-1.29}^{+1.08}$ & {\color{red}$0.60\pm 0.01$} & {\color{red}$0.28\pm 0.02$} & {\color{red}$0.41\pm 0.01$} & {\color{red}$1.67_{-0.06}^{+0.05}$} & {\color{red}$19.1_{-0.8}^{+0.7}$} & {\color{red}$2.94_{-0.22}^{+0.19}$} \\
NGC4631 & $18.55_{-0.69}^{+0.66}$ & {\color{red}$0.35\pm 0.01$} & {\color{red}$0.47_{-0.01}^{+0.03}$} & {\color{red}$0.95_{-0.01}^{+0.03}$} & {\color{red}$1.21_{-0.02}^{+0.03}$} & {\color{red}$8.0_{-0.2}^{+0.3}$} & {\color{red}$1.36_{-0.06}^{+0.07}$} \\
NGC4666 & $27.01_{-11.97}^{+4.41}$ & $0.27_{-0.05}^{+0.04}$ & {\color{red}$0.69_{-0.29}^{+0.12}$} & {\color{red}$1.77_{-0.37}^{+0.15}$} & {\color{red}$0.97_{-0.20}^{+0.08}$} & {\color{red}$5.0_{-1.4}^{+0.8}$} & {\color{red}$0.58_{-0.31}^{+0.14}$} \\
NGC4710 & $6.00_{-3.51}^{+0.80}$ & $0.63_{-0.06}^{+0.10}$ & {\color{red}$0.080_{-0.038}^{+0.013}$} & {\color{red}$1.46_{-0.34}^{+0.11}$} & {\color{red}$0.14_{-0.03}^{+0.01}$} & {\color{red}$1.6_{-0.4}^{+0.3}$} & {\color{red}$0.85_{-0.54}^{+0.19}$} \\
NGC5102 & {\color{red}$0.58\pm 0.11$} & 0.3 & {\color{red}$0.013\pm 0.002$} & {\color{red}$3.46\pm 0.34$} & {\color{red}$0.0091\pm 0.0009$} & {\color{red}$0.052\pm 0.018$} & {\color{red}$0.28\pm 0.11$} \\
NGC5170 & {\color{red}$24.85\pm 6.30$} & 0.3 & {\color{red}$0.90\pm 0.23$} & {\color{red}$0.74\pm 0.09$} & {\color{red}$2.99\pm 0.38$} & {\color{red}$17.1\pm 6.1$} & {\color{red}$2.18\pm 0.95$} \\
NGC5253 & {\color{red}$1.09\pm 0.07$} & $0.35_{-0.01}^{+0.02}$ & {\color{red}$0.028\pm 0.002$} & {\color{red}$2.55_{-0.09}^{+0.08}$} & {\color{red}$0.0270_{-0.0009}^{+0.0008}$} & {\color{red}$0.179_{-0.008}^{+0.012}$} & {\color{red}$0.52_{-0.04}^{+0.05}$} \\
NGC5422 & {\color{red}$18.36\pm 3.36$} & 0.3 & {\color{red}$0.35\pm 0.06$} & {\color{red}$1.74\pm 0.16$} & {\color{red}$0.49\pm 0.05$} & {\color{red}$2.8\pm 1.0$} & {\color{red}$0.48\pm 0.19$} \\
NGC5746 & $12.87_{-7.50}^{+4.79}$ & $0.16_{-0.16}^{+0.12}$ & {\color{red}$0.46_{-0.14}^{+9.57}$} & {\color{red}$0.84_{-0.13}^{+8.75}$} & {\color{red}$1.35_{-0.21}^{+14.07}$} & {\color{red}$4.2_{-4.2}^{+43.7}$} & {\color{red}$1.03_{-1.20}^{+10.77}$} \\
NGC5775 & $36.35_{-4.25}^{+3.64}$ & $0.38_{-0.04}^{+0.05}$ & {\color{red}$0.92_{-0.19}^{+0.11}$} & {\color{red}$0.79_{-0.08}^{+0.05}$} & {\color{red}$2.87_{-0.30}^{+0.17}$} & {\color{red}$20.9_{-3.0}^{+2.8}$} & {\color{red}$1.82_{-0.34}^{+0.30}$} \\
NGC5866 & $8.96_{-1.56}^{+1.34}$ & $0.31_{-0.03}^{+0.04}$ & {\color{red}$0.15\pm 0.03$} & {\color{red}$1.00_{-0.10}^{+0.11}$} & {\color{red}$0.38\pm 0.04$} & {\color{red}$2.3_{-0.3}^{+0.4}$} & {\color{red}$0.81\pm 0.18$} \\
NGC6503 & $1.54_{-0.21}^{+0.18}$ & $0.42_{-0.06}^{+0.09}$ & {\color{red}$0.039\pm 0.005$} & {\color{red}$2.36\pm 0.15$} & {\color{red}$0.041\pm 0.003$} & {\color{red}$0.32_{-0.05}^{+0.07}$} & {\color{red}$0.67_{-0.14}^{+0.17}$} \\
NGC6764 & $23.48_{-9.84}^{+6.76}$ & $0.75_{-0.11}^{+0.13}$ & {\color{red}$0.70_{-0.26}^{+0.17}$} & {\color{red}$4.47_{-0.83}^{+0.55}$} & {\color{red}$0.38_{-0.07}^{+0.05}$} & {\color{red}$5.5_{-1.3}^{+1.2}$} & {\color{red}$0.74_{-0.36}^{+0.27}$} \\
NGC7090 & $0.44_{-0.38}^{+0.32}$ & $0.44_{-0.14}^{+0.13}$ & {\color{red}$0.009_{-0.005}^{+0.011}$} & {\color{red}$1.52_{-0.41}^{+0.89}$} & {\color{red}$0.015_{-0.004}^{+0.009}$} & {\color{red}$0.13_{-0.05}^{+0.08}$} & {\color{red}$0.91_{-0.87}^{+0.89}$} \\
NGC7457 & {\color{red}$4.97\pm 1.26$} & 0.3 & {\color{red}$0.12\pm 0.03$} & {\color{red}$1.34\pm 0.17$} & {\color{red}$0.21\pm 0.03$} & {\color{red}$1.2\pm 0.4$} & {\color{red}$0.78\pm 0.34$} \\
NGC7582 & $62.94_{-11.88}^{+10.51}$ & $0.67_{-0.07}^{+0.08}$ & {\color{red}$1.64_{-0.27}^{+0.38}$} & {\color{red}$2.60_{-0.21}^{+0.30}$} & {\color{red}$1.55_{-0.13}^{+0.18}$} & {\color{red}$19.7_{-2.7}^{+3.2}$} & {\color{red}$1.00\pm 0.23$} \\
NGC7814 & {\color{red}$5.89\pm 0.97$} & 0.3 & {\color{red}$0.18\pm 0.03$} & {\color{red}$3.88\pm 0.32$} & {\color{red}$0.116\pm 0.009$} & {\color{red}$0.66\pm 0.23$} & {\color{red}$0.36\pm 0.14$} \\
\enddata
\tablecomments{\scriptsize Hot gas properties: (1) absorption corrected 0.5-2~keV luminosity; (2) temperature; (3) volume emission measure; (4) electron number density; (5) mass; (6) thermal energy; (7) radiative cooling timescale. All the parameters are estimated from the 1-T fit with fixed abundance ratio. The temperature is fixed at $0.3\rm~keV$ for some galaxies with low counting statistic (those without errors). See 
\S3.3
for details.}\label{L13atable:1Tspec}
\end{deluxetable}

\begin{deluxetable}{lcccccccc}[!h]
\centering
\tiny 
  \tabletypesize{\tiny}
  \tablecaption{Hot Gas Properties from 1-T Model Fit with Free Abundance Ratio (Table~8 of \citet{Li13a})}
  \tablewidth{0pt}
  \tablehead{
 \colhead{Name} & \colhead{$L_X$} & \colhead{$T_X$} & \colhead{$EM$} & \colhead{$n_e$} & \colhead{$M_{hot}$} & \colhead{$E_{hot}$} & \colhead{$t_{cool}$} & \colhead{$Fe/\alpha$}\\
    & ($10^{38}\rm ergs/s$) & (keV) & ($10^{-2}\rm cm^{-6}kpc^3$) & ($f^{-1/2}10^{-3}\rm cm^{-3}$) & ($f^{1/2}10^8\rm M_\odot$) & ($f^{1/2}10^{55}\rm ergs$) & ($f^{1/2}\rm Gyr$) & \\
   & (1) & (2) & (3) & (4) & (5) & (6) & (7) & (8)
}
\startdata
M82 & $67.68_{-0.38}^{+0.39}$ & {\color{red}$0.616\pm 0.003$} & {\color{red}$1.63_{-0.01}^{+0.02}$} & {\color{red}$9.04_{-0.03}^{+0.07}$} & {\color{red}$0.445_{-0.002}^{+0.003}$} & {\color{red}$5.22_{-0.03}^{+0.05}$} & {\color{red}$0.245_{-0.002}^{+0.003}$} & {\color{red}$0.36\pm 0.01$} \\
NGC0891 & $21.85_{-0.60}^{+0.48}$ & {\color{red}$0.32\pm 0.01$} & {\color{red}$0.51_{-0.02}^{+0.01}$} & {\color{red}$1.20\pm 0.02$} & {\color{red}$1.05\pm 0.02$} & {\color{red}$6.4\pm 0.2$} & {\color{red}$0.93_{-0.04}^{+0.03}$} & $0.43_{-0.04}^{+0.06}$ \\
NGC1482 & $48.96_{-2.73}^{+3.57}$ & $0.49_{-0.04}^{+0.03}$ & {\color{red}$1.12_{-0.19}^{+0.21}$} & {\color{red}$5.73_{-0.48}^{+0.53}$} & {\color{red}$0.48\pm 0.04$} & {\color{red}$4.6\pm 0.5$} & {\color{red}$0.30\pm 0.04$} & $0.36_{-0.06}^{+0.11}$ \\
NGC1808 & $16.60_{-0.79}^{+0.73}$ & {\color{red}$0.51\pm 0.03$} & {\color{red}$0.40\pm 0.04$} & {\color{red}$3.23\pm 0.16$} & {\color{red}$0.31_{-0.01}^{+0.02}$} & {\color{red}$2.9\pm 0.2$} & {\color{red}$0.56\pm 0.05$} & $0.34_{-0.05}^{+0.06}$ \\
NGC2841 & $17.75_{-2.83}^{+2.48}$ & $0.38_{-0.05}^{+0.04}$ & {\color{red}$0.39_{-0.04}^{+0.08}$} & {\color{red}$1.39_{-0.07}^{+0.14}$} & {\color{red}$0.69_{-0.04}^{+0.07}$} & {\color{red}$5.0\pm 0.7$} & {\color{red}$0.89_{-0.19}^{+0.18}$} & $0.48_{-0.12}^{+0.19}$ \\
NGC3079 & $65.29_{-4.96}^{+4.99}$ & {\color{red}$0.51\pm 0.03$} & {\color{red}$1.65\pm 0.14$} & {\color{red}$1.53_{-0.06}^{+0.07}$} & {\color{red}$2.67\pm 0.11$} & {\color{red}$25.8_{-1.8}^{+1.9}$} & {\color{red}$1.25\pm 0.13$} & $0.31_{-0.05}^{+0.06}$ \\
NGC3556 & $5.73_{-1.18}^{+1.03}$ & $0.33_{-0.03}^{+0.06}$ & {\color{red}$0.14_{-0.02}^{+0.03}$} & {\color{red}$0.80_{-0.07}^{+0.09}$} & {\color{red}$0.45_{-0.04}^{+0.05}$} & {\color{red}$2.8_{-0.4}^{+0.6}$} & {\color{red}$1.55_{-0.38}^{+0.42}$} & {\color{red}$0.30\pm 0.12$} \\
NGC3628 & $23.32_{-2.08}^{+1.08}$ & $0.30_{-0.02}^{+0.01}$ & {\color{red}$0.47_{-0.04}^{+0.05}$} & {\color{red}$0.66_{-0.03}^{+0.04}$} & {\color{red}$1.77_{-0.08}^{+0.09}$} & {\color{red}$10.2_{-0.8}^{+0.7}$} & {\color{red}$1.39_{-0.16}^{+0.11}$} & $0.45_{-0.08}^{+0.48}$ \\
NGC4342 & $91.18_{-3.34}^{+3.53}$ & {\color{red}$0.54\pm 0.02$} & {\color{red}$1.40_{-0.20}^{+0.21}$} & {\color{red}$1.35\pm 0.10$} & {\color{red}$2.56_{-0.18}^{+0.20}$} & {\color{red}$26.3_{-2.0}^{+2.2}$} & {\color{red}$0.92\pm 0.08$} & $0.79_{-0.12}^{+0.16}$ \\
NGC4526 & $8.77_{-1.95}^{+1.77}$ & $0.26_{-0.04}^{+0.07}$ & {\color{red}$0.16\pm 0.04$} & {\color{red}$0.93_{-0.11}^{+0.13}$} & {\color{red}$0.42_{-0.05}^{+0.06}$} & {\color{red}$2.1_{-0.4}^{+0.6}$} & {\color{red}$0.75_{-0.22}^{+0.26}$} & $1.25_{-0.78}^{+1.25}$ \\
NGC4594 & $21.95_{-1.77}^{+1.66}$ & {\color{red}$0.60\pm 0.01$} & {\color{red}$0.33\pm 0.08$} & {\color{red}$0.45_{-0.06}^{+0.05}$} & {\color{red}$1.83_{-0.23}^{+0.21}$} & {\color{red}$20.9_{-2.7}^{+2.5}$} & {\color{red}$3.03_{-0.46}^{+0.42}$} & $0.81_{-0.19}^{+0.29}$ \\
NGC4631 & $18.16_{-0.75}^{+0.72}$ & {\color{red}$0.33\pm 0.01$} & {\color{red}$0.44\pm 0.02$} & {\color{red}$0.93\pm 0.02$} & {\color{red}$1.18_{-0.03}^{+0.02}$} & {\color{red}$7.5_{-0.3}^{+0.2}$} & {\color{red}$1.31\pm 0.07$} & {\color{red}$0.37\pm 0.04$} \\
NGC5253 & {\color{red}$1.04\pm 0.08$} & {\color{red}$0.32\pm 0.02$} & {\color{red}$0.024\pm 0.002$} & {\color{red}$2.34\pm 0.10$} & {\color{red}$0.025\pm 0.001$} & {\color{red}$0.15\pm 0.01$} & {\color{red}$0.46\pm 0.05$} & $0.48_{-0.07}^{+0.08}$ \\
NGC5866 & $10.59_{-2.12}^{+1.43}$ & $0.137_{-0.001}^{+0.004}$ & {\color{red}$0.17_{-0.09}^{+0.05}$} & {\color{red}$1.05_{-0.27}^{+0.14}$} & {\color{red}$0.40_{-0.10}^{+0.06}$} & {\color{red}$1.1_{-0.3}^{+0.1}$} & {\color{red}$0.32_{-0.10}^{+0.06}$} & $<401.10$ \\
\enddata
\tablecomments{\scriptsize Hot gas properties of the 1-T fit with the Fe/$\alpha$ ratio set free: (1) extinction corrected 0.5-2~keV luminosity; (2) temperature; (3) volume emission measure; (4) electron number density; (5) mass; (6) thermal energy; (7) radiative cooling timescale; (8) Fe/$\alpha$ ratio. See 
\S3.3 for details.}\label{Li13atable:specpara1Tmetal}
\end{deluxetable}

\begin{deluxetable}{lcccccccc}[!h]
\centering
\tiny 
  \tabletypesize{\tiny}
  \tablecaption{Hot Gas Properties from the 2-T Plasma Model (Table~9 of \citet{Li13a})}
  \tablewidth{0pt}
  \tablehead{
 \colhead{Name} & \colhead{$L_{X,low}$} & \colhead{$T_{X,low}$} & \colhead{$EM_{X,Low}$} & \colhead{$L_{X,high}$} & \colhead{$T_{X,high}$} & \colhead{$EM_{X,high}$} & \colhead{$L_{X,total}$} & \colhead{$T_{X,L_W}$} \\
    & ($10^{38}\rm ergs/s$) & (keV) & ($10^{-2}\rm cm^{-6}kpc^3$) & ($10^{38}\rm ergs/s$) & (keV) & ($10^{-2}\rm cm^{-6}kpc^3$) & ($10^{38}\rm ergs/s$) & (keV) \\
    & (1) & (2) & (3) & (4) & (5) & (6) & (7) & (8)
}
\startdata
NGC0520 & $23.08_{-20.57}^{+7.05}$ & $0.35_{-0.09}^{+0.07}$ & {\color{red}$0.59_{-0.28}^{+0.22}$} & $12.94(<22.76)$ & $0.77_{-0.19}^{+0.27}$ & {\color{red}$0.36_{-0.26}^{+0.49}$} & $36.1_{-6.9}^{+4.5}$ & $0.50_{-0.13}^{+0.14}$ \\
NGC2841 & $4.20_{-1.57}^{+1.40}$ & $0.11_{-0.02}^{+0.03}$ & {\color{red}$0.62_{-0.24}^{+0.79}$} & $23.56_{-1.76}^{+2.31}$ & $0.49_{-0.05}^{+0.04}$ & {\color{red}$0.60_{-0.04}^{+0.06}$} & $27.7_{-1.9}^{+3.0}$ & $0.43_{-0.05}^{+0.04}$ \\
NGC3079 & $10.28_{-1.53}^{+1.49}$ & $0.12_{-0.01}^{+0.02}$ & {\color{red}$0.90\pm 0.26$} & $77.08_{-3.52}^{+2.64}$ & {\color{red}$0.56\pm 0.02$} & {\color{red}$1.96_{-0.09}^{+0.08}$} & $87.1_{-3.6}^{+2.9}$ & {\color{red}$0.51\pm 0.02$} \\
NGC3521 & $14.00(<19.75)$ & {\color{red}$0.31_{-0.05}^{+0.00}$} & {\color{red}$0.35_{-0.15}^{+0.38}$} & $7.18(<14.12)$ & $0.56(<4.09)$ & {\color{red}$0.19(<0.33)$} & $21.2_{-2.0}^{+1.7}$ & $0.40_{-0.22}^{+1.20}$ \\
NGC3556 & $3.29_{-0.66}^{+0.57}$ & {\color{red}$0.20\pm 0.03$} & {\color{red}$0.096\pm 0.014$} & $5.94_{-0.76}^{+0.58}$ & $0.61_{-0.03}^{+0.05}$ & {\color{red}$0.15_{-0.01}^{+0.02}$} & {\color{red}$9.2\pm 0.5$} & $0.47_{-0.03}^{+0.05}$ \\
NGC3628 & $7.37_{-1.71}^{+5.42}$ & $0.14_{-0.02}^{+0.03}$ & {\color{red}$0.48_{-0.13}^{+0.17}$} & $25.01_{-6.76}^{+2.23}$ & $0.40_{-0.03}^{+0.19}$ & {\color{red}$0.64_{-0.07}^{+0.06}$} & $34.0_{-3.8}^{+2.5}$ & $0.34_{-0.03}^{+0.15}$ \\
NGC3877 & $0.80(<1.13)$ & $0.18(<0.24)$ & {\color{red}$0.027_{-0.015}^{+0.231}$} & $1.38_{-0.62}^{+0.28}$ & $0.59_{-0.24}^{+0.13}$ & {\color{red}$0.035_{-0.014}^{+0.009}$} & $2.1_{-0.8}^{+0.3}$ & $0.44_{-0.21}^{+0.11}$ \\
NGC4388 & $5.61_{-2.53}^{+1.81}$ & $0.09(<0.12)$ & {\color{red}$1.53_{-0.87}^{+1.76}$} & $51.82_{-3.31}^{+3.09}$ & {\color{red}$0.61\pm 0.04$} & {\color{red}$1.34_{-0.14}^{+0.10}$} & $57.4_{-7.1}^{+3.1}$ & $0.56_{-0.05}^{+0.04}$ \\
NGC4501 & $3.06(<10.59)$ & $0.29(<0.29)$ & {\color{red}$0.100(<0.368)$} & $30.48(<37.90)$ & {\color{red}$0.58_{-0.08}^{+0.00}$} & {\color{red}$0.75_{-0.19}^{+0.26}$} & $33.6_{-6.9}^{+8.1}$ & {\color{red}$0.55_{-0.10}^{+0.00}$} \\
NGC4565 & $4.98(<8.70)$ & {\color{red}$0.29_{-0.10}^{+0.00}$} & {\color{red}$0.13_{-0.12}^{+0.11}$} & $4.74(<6.62)$ & $0.56(<0.56)$ & {\color{red}$0.12(<0.16)$} & $9.7_{-1.3}^{+1.0}$ & {\color{red}$0.42_{-0.32}^{+0.00}$} \\
NGC4569 & $4.19_{-1.51}^{+0.88}$ & $0.10(<0.15)$ & {\color{red}$2.48_{-1.94}^{+0.63}$} & $13.22_{-1.92}^{+1.13}$ & $0.56_{-0.05}^{+0.02}$ & {\color{red}$0.34\pm 0.03$} & $17.4_{-2.8}^{+1.0}$ & $0.45_{-0.06}^{+0.03}$ \\
NGC4631 & $9.33_{-1.15}^{+0.97}$ & $0.24_{-0.02}^{+0.01}$ & {\color{red}$0.24_{-0.03}^{+0.02}$} & $13.77_{-1.41}^{+1.53}$ & {\color{red}$0.58\pm 0.02$} & {\color{red}$0.36\pm 0.04$} & $23.3_{-1.1}^{+0.9}$ & {\color{red}$0.44\pm 0.02$} \\
NGC5253 & $0.20_{-0.06}^{+0.08}$ & $0.14_{-0.02}^{+0.08}$ & {\color{red}$0.012_{-0.003}^{+0.006}$} & $1.04_{-0.09}^{+0.07}$ & $0.40_{-0.02}^{+0.03}$ & {\color{red}$0.026_{-0.002}^{+0.001}$} & {\color{red}$1.24\pm 0.04$} & $0.36_{-0.02}^{+0.04}$ \\
NGC5775 & $6.85_{-3.81}^{+9.90}$ & $0.08(<0.20)$ & {\color{red}$1.19_{-0.99}^{+15.78}$} & $38.22(<40.85)$ & $0.37(<0.61)$ & {\color{red}$0.96_{-0.71}^{+0.08}$} & $45.1_{-3.6}^{+3.1}$ & $0.33(<0.55)$ \\
NGC6764 & $4.59(<10.84)$ & $0.30(<0.30)$ & {\color{red}$0.13(<0.85)$} & $24.36_{-19.57}^{+5.84}$ & {\color{red}$0.75_{-0.10}^{+0.00}$} & {\color{red}$0.69(<0.87)$} & $29.6_{-11.1}^{+9.8}$ & {\color{red}$0.68_{-0.13}^{+0.00}$} \\
NGC7582 & $14.61(<17.97)$ & $0.09(<0.11)$ & {\color{red}$4.14_{-3.51}^{+1.95}$} & $90.19_{-28.83}^{+12.12}$ & {\color{red}$0.64\pm 0.05$} & {\color{red}$2.43_{-0.32}^{+0.33}$} & $105.0_{-6.8}^{+11.6}$ & $0.56_{-0.05}^{+0.04}$ \\
\enddata
\tablecomments{\scriptsize See 
\S3.3 for details.}\label{Li13atable:specpara2T}
\end{deluxetable}

\section{Erratum: ``D\lowercase{o We Detect the Galactic Feedback Material in} X-\lowercase{ray Observations of Nearby Galaxies?} - A \lowercase{Case Study of} NGC 5866" (\href{https://doi.org/10.1093/mnras/stv1701}{2015, MNRAS, 453, 1062})} \label{sec:Li15a}

$\bullet$ EM in Table~1 of \citet{Li15a} should be updated to those in Table~\ref{Li15atable:SpecFitResult} of this erratum.

$\bullet$ Section~3.1, 2nd paragraph. The following sentences should be changed to:

``The densities of the inner and outer halos are {\color{red}$\sim3.8\times10^{-3}f^{-1/2}\rm~cm^{-3}$} and {\color{red}$\sim1.1\times10^{-3}f^{-1/2}\rm~cm^{-3}$}, respectively, where $f$ is the poorly known volume filling factor. The total mass of hot gas contained in the inner and outer halos are both {\color{red}$\sim3\times10^7f^{1/2}\rm~M_\odot$}. As the metallicity in the outer halo cannot be well constrained, we only estimate the total Fe mass contained in the inner halo, which is {\color{red}$\sim6\times10^4f^{1/2}\rm~M_\odot$}.''

$\bullet$ Section~3.1, 3rd paragraph. The following sentences should be changed to:

``Within this timescale, there are $\sim2\times10^4\rm~M_\odot$ Fe injected into the ISM by Type~Ia SNe (assuming each Type~Ia SN eject $\sim0.7\rm~M_\odot$ Fe; Nomoto, Thielemann \& Yokoi 1984), about {\color{red}three times} smaller than what we have detected in the inner halo, assuming the volume filling factor $f\sim1$.''

``Within this timescale, the total Fe injected by Type~Ia SNe is $\sim10^7\rm~M_\odot$, {\color{red}more than two orders of magnitude} higher than what we have detected.''

\begin{table}[!h]{}
\vspace{-0.in}
\begin{center}
\small\caption{(Table~1 of \citet{Li15a}) Hot gas properties from spectral analysis. Spectral analysis results of the `inner halo' region is obtained by jointly fitting the \emph{Suzaku} and \emph{Chandra} spectra (in total 10 free parameters). Spectral analysis results of the `outer halo' region is obtained by fitting the \emph{Suzaku} XIS1 spectrum with a model in which the absorption column density and photon index of the power law, as well as the O and Fe abundances of the hot gas, are all fixed at the values of the inner halo (in total five free parameters). See text for the detail description of the spectral models and Figs~2(a) and (c) for the fitted spectra. All the errors quoted in this table are statistical only, at 90 per cent confidence level. EM is the emission measure of the VAPEC component.}
\begin{tabular}{lcc}
\hline \hline
Parameter        & Inner halo      & Outer halo \\
\hline
$\chi^2/d.o.f.$  & $121.00/135$ & $6.59/15$ \\
$kT\rm~(keV)$ & $0.248_{-0.027}^{+0.026}$ & $0.66_{-0.11}^{+0.16}$ \\
O abundance (solar) & $0.21_{-0.10}^{+0.13}$ & - \\
Fe abundance (solar) & $1.57_{-0.58}^{+1.32}$ & - \\
EM ($\rm cm^{-6}kpc^3$) & {\color{red}$0.0053_{-0.0012}^{+0.0015}$} & {\color{red}$0.0015\pm0.0004$} \\
\hline \hline
\end{tabular}\label{Li15atable:SpecFitResult}
\end{center}
\end{table}

\section{Erratum: ``XMM-N\lowercase{ewton Large Program on} SN1006 - I: M\lowercase{ethods and Initial Results of Spatially-Resolved Spectroscopy}" (\href{https://doi.org/10.1093/mnras/stv1882}{2015, MNRAS, 453, 3953})} \label{sec:Li15b}

\begin{figure}[!h]
\begin{center}
\epsfig{figure=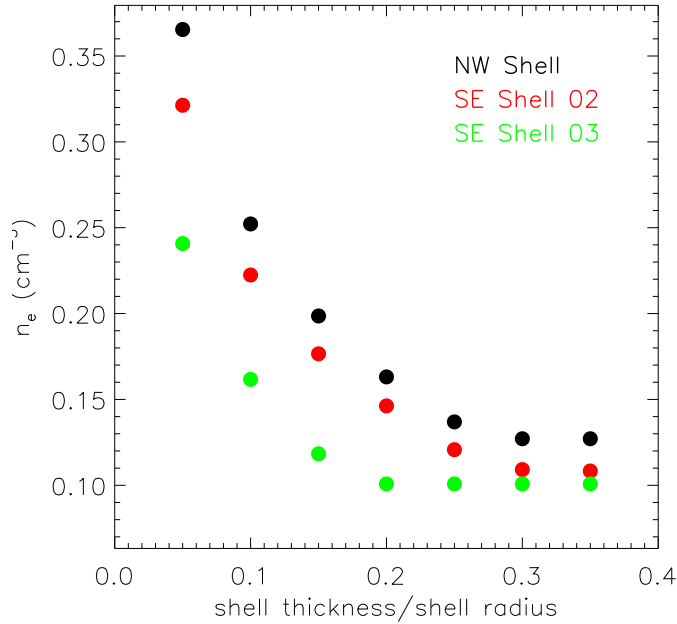,width=0.5\textwidth,angle=0, clip=}
\caption{(Fig.~5 of \citet{Li15b}) Derived post-shock electron number density in several outer regions vs. the assumed thickness of the thermal X-ray emitting shell (in unit of the outer radius of the shell). Region names are denoted in Fig.~7.}\label{Li15bPaperIfig:ShellThickness}
\end{center}
\end{figure}

$\bullet$ Fig.~5 of \citet{Li15b} should be updated to Fig.~\ref{Li15bPaperIfig:ShellThickness} of this erratum. Only the scale of the vertical axis has been changed.

$\bullet$ $n_e$ in Table~4 of \citet{Li15b} should be updated to those in Table~\ref{Li15bPaperItable:ParaIndividualSpec} of this erratum.

\begin{figure}[!h]
\begin{center}
\epsfig{figure=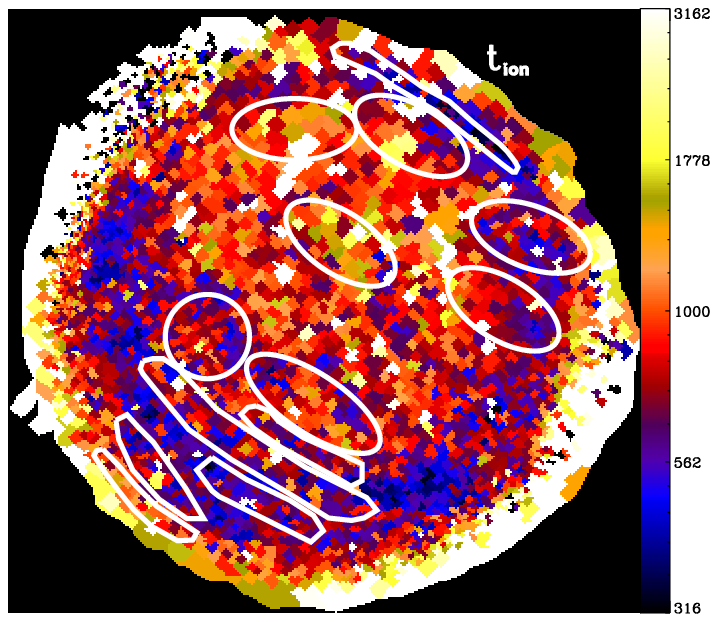,height=0.33\textwidth,angle=0, clip=}
\epsfig{figure=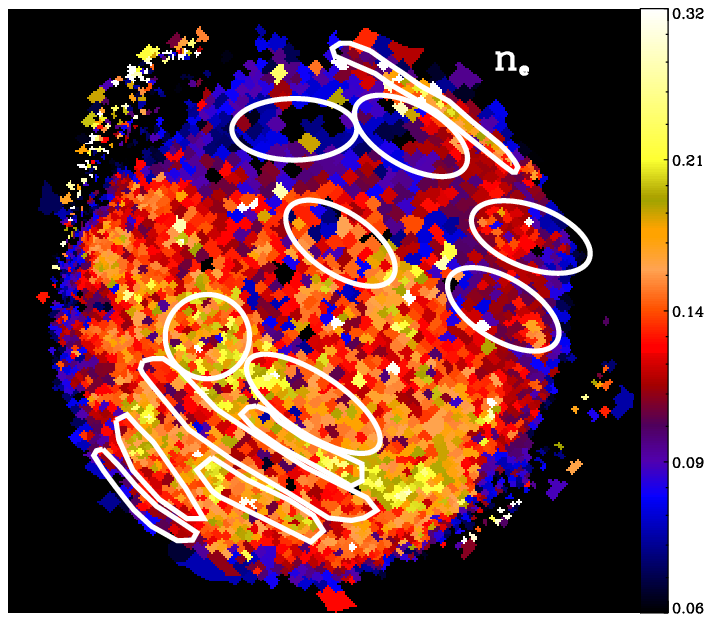,height=0.33\textwidth,angle=0, clip=}
\caption{Updated version of Fig.~9(c) and (d) of \citet{Li15b}.}\label{Li15bPaperIfig:2DSpec_paraimg}
\end{center}
\end{figure}

$\bullet$ Fig.~9(c) and (d) of \citet{Li15b} should be updated to the left and right panels of Fig.~\ref{Li15bPaperIfig:2DSpec_paraimg} of this erratum, respectively. While the $n_{\rm e}$ map (right panel) only has the scale of the color bar changed, the $t_{\rm ion}$ map (left panel) is further slightly changed compared to the original version, assuming a shell thickness of 0.2 shell radius, consistent with other maps in Fig.~9 (see section 3.2.2 of the text for details).

$\bullet$ The following sentence in the 2nd from the last paragraph of section~3.2.2 should be changed to:

``As will be discussed in Section~4.2.1, the maximum value of $t_{ion}$ in the SNR interior is typically {\color{red}$\sim1500\rm~yr$}, consistent with the age of SN1006 ($\sim10^3\rm~yr$ based on historical records; Stephenson 2010).''

$\bullet$ The following sentences in the last paragraph of section~4.2.1 should be changed to:

``Except for the bright rim surrounding the SNR, which is artificial due to the low flux density of the surrounding regions, the whole SNR shell appears to have a low and smooth ionization age of {\color{red}$t_{ion}\lesssim500\rm~year$}. In contrast, all the regions in the SNR interior have {\color{red}$t_{ion}>500\rm~year$}, with the highest {\color{red}$t_{ion}\sim1500\rm~year$}, consistent with the age of SN1006 of $\sim10^3\rm~year$.''

$\bullet$ The following sentences in section~4.2.2 should be changed to:

``Assuming no CR acceleration in this part so a compression ratio of 4, the ambient ISM density surrounding the ``NW shell'' should be {\color{red}$n_0\sim0.05\rm~cm^{-3}$, significantly lower than the values from previous multi-wavelength estimates ($n_0\sim0.4\rm~cm^{-3}$), which probably indicates the thickness of the SNR shell (0.2 times of the SNR radius) has been overestimated.}'' 

``The northern part of the ``SNR Interior'' has a clearly lower density of {\color{red}$n_e\lesssim0.1\rm~cm^{-3}$}, but ``SNR Interior 03 and 04'' may form a shell-like structure behind the SW non-thermal rim, apparently extending the ``NW Shell''.''

``The estimated mass of the shocked X-ray emitting plasma is {\color{red}$\sim5f^{\frac{1}{2}}\rm~M_\odot$}, where $f$ is the volume filling factor. As discussed in Section~3.2.2, the swept up ambient ISM mass is $\sim5\rm~M_\odot$. Adding the mass of the shocked ejecta, which is quite uncertain but contribute only a small fraction to the mass budget, we could roughly constrain the volume filling factor to be {\color{red}$f\sim1$} under the adopted geometric model (Section~3.2.2).''

\begin{deluxetable}{llccccccccccc}
\centering
\scriptsize 
  \tabletypesize{\scriptsize}
  \tablecaption{Average value of parameters for individual regions (Table~4 of \citet{Li15b})}
  \tablehead{
  \colhead{Region} & \colhead{Method} & \colhead{$kT$} & \colhead{$n_e$} & \colhead{$\log n_et$} & \colhead{\ion{O}{7}~$\rm K\delta-\zeta$} & \colhead{\ion{O}{7}} & \colhead{\ion{O}{8}} & \colhead{O} & \colhead{Ne} & \colhead{Mg} & \colhead{Si} & \colhead{Fe} \\
    & & keV & $\rm cm^{-3}$ & $\log(\rm cm^{-3}s)$ & EW & EW & EW & solar & solar & solar & solar & solar 
 }
\startdata
NW Shell           & Average      	& 2.25 & {\color{red}0.16} & 9.56 & 0.18, 23.3 & 0.52, 12.0 & 0.33, 51.9 & 1.13 & 0.85 & 2.20 & 11.06 & 0.23 \\
		        & Fit		& 1.58 & {\color{red}0.12} & 9.32 & -                & -                & -               & 0.92 & 0.64 & 1.05 & 4.84 & 0.05 \\
\hline
SNR Interior 01 & Average      	& 2.10 & {\color{red}0.082} & 9.50 & 0.15, 33.1 & 0.57, 15.3 & 0.33, 61.9 & 1.77 & 0.49 & 3.81 & 32.38 & 0.18 \\
		        & Fit		& 2.22 & {\color{red}0.076} & 9.37 & -                & -                & -               & 0.90 & 0.27 & 1.14 & 7.85 & 0.39 \\
\hline
SNR Interior 02 & Average      	& 1.63 & {\color{red}0.104} & 9.51 & 0.21, 38.3 & 0.64, 17.8 & 0.41, 74.9 & 2.07 & 0.98 & 5.11 & 30.71 & 0.52 \\
		        & Fit		& 1.56 & {\color{red}0.098} & 9.35 & -                & -                & -               & 1.05 & 0.44 & 1.69 & 12.36 & 0.83 \\
\hline
SNR Interior 03 & Average      	& 2.47 & {\color{red}0.108} & 9.40 & 0.17, 23.8 & 0.60, 13.2 & 0.36, 51.5 & 1.44 & 0.50 & 3.42 & 6.95 & 0.25 \\
		        & Fit		& 4.59 & {\color{red}0.089} & 9.35 & -                & -                & -               & 1.23 & 0.32 & 1.46 & 5.31 & 0.05 \\
\hline
SNR Interior 04 & Average      	& 2.87 & {\color{red}0.12} & 9.48 & 0.19, 27.0 & 0.60, 10.6 & 0.44, 52.3 & 1.26 & 0.36 & 1.97 & 9.83 & 0.40 \\
		        & Fit		& 2.30 & {\color{red}0.11} & 9.40 & -                & -                & -               & 1.01 & 0.28 & 1.53 & 7.36 & 0.36 \\
\hline
SNR Interior 05 & Average      	& 1.30 & {\color{red}0.13} & 9.65 & 0.21, 34.8 & 0.43, 10.8 & 0.37, 59.3 & 1.53 & 0.62 & 3.16 & 20.23 & 0.18 \\
		        & Fit		& 1.12 & {\color{red}0.11} & 9.60 & -                & -                & -               & 0.80 & 0.35 & 1.26 & 9.31 & 0.28 \\
\hline
Dark Belt           & Average      	& 2.06 & {\color{red}0.14} & 9.49 & 0.17, 25.1 & 0.46, 8.99 & 0.33, 44.7 & 1.06 & 0.39 & 1.10 & 15.33 & 0.73 \\
		        & Fit		& 2.54 & {\color{red}0.10} & 9.41 & -                & -                & -               & 0.87 & 0.32 & 0.94 & 9.30 & 0.68 \\
\hline
Interior Shell 01 & Average      	& 2.41 & {\color{red}0.17} & 9.58 & 0.22, 35.9 & 0.40, 8.43 & 0.37, 50.5 & 1.06 & 0.42 & 0.86 & 9.61 & 0.99 \\
		        & Fit		& 2.20 & {\color{red}0.14} & 9.53 & -                & -                & -               & 0.97 & 0.38 & 0.82 & 8.86 & 0.96 \\
\hline
Interior Shell 02 & Average      	& 2.36 & {\color{red}0.17} & 9.59 & 0.20, 31.1 & 0.37, 8.85 & 0.37, 53.0 & 1.23 & 0.35 & 1.62 & 13.41 & 0.34 \\
		        & Fit		& 1.61 & {\color{red}0.16} & 9.54 & -                & -                & -               & 0.77 & 0.26 & 1.01 & 8.67 & 0.44 \\
\hline
O hole               & Average      	& 1.70 & {\color{red}0.16} & 9.61 & 0.16, 31.1 & 0.28, 5.74 & 0.27, 38.7 & 0.81 & 0.30 & 0.84 & 13.93 & 1.04 \\
		        & Fit		& 1.65 & {\color{red}0.17} & 9.61 & -                & -                & -               & 0.43 & 0.21 & 0.47 & 5.48 & 0.63 \\
\hline
SE Shell 01       & Average      	& 1.89 & {\color{red}0.17} & 9.53 & 0.17, 28.3 & 0.40, 8.38 & 0.37, 47.5 & 1.10 & 0.27 & 1.44 & 16.56 & 0.69 \\
		        & Fit		& 2.32 & {\color{red}0.13} & 9.45 & -                & -                & -               & 0.99 & 0.24 & 1.32 & 12.90 & 0.76 \\
\hline
SE Shell 02       & Average      	& 1.39 & {\color{red}0.15} & 9.52 & 0.16, 27.5 & 0.51, 10.0 & 0.39, 53.2 & 1.33 & 0.34 & 2.65 & 22.06 & 0.28 \\
		        & Fit		& 1.07 & {\color{red}0.13} & 9.58 & -                & -                & -               & 1.23 & 0.31 & 1.82 & 17.61 & 0.18 \\
\hline
SE Shell 03       & Average      	& 1.65 & {\color{red}0.101} & 9.54 & 0.22, 27.4 & 0.64, 9.63 & 0.60, 55.0 & 1.45 & 0.28 & 3.21 & 15.85 & 0.10 \\
		        & Fit		& 1.66 & {\color{red}0.095} & 9.48 & -                & -                & -               & 1.15 & 0.22 & 2.24 & 11.52 & 0.05 
\enddata
\tablecomments{\scriptsize Average parameters of large regions enclosing some interesting features as denoted in Fig.~7. For each region, the average parameters are calculated in two ways: a direct average based on the parameter images (``Average'') and the parameters obtained by fitting the MOS-1+MOS-2+PN spectra extracted from each region (e.g., Fig.~8) using the model described in Section~3.2.1 (``Fit''). For the former method, $kT$, $\log n_et$, and $n_e$ are calculated from Fig.~9(a), (b), and (d). \ion{O}{7}, \ion{O}{8}, and \ion{O}{7}~$\rm K\delta-\zeta$ EWs are calculated from the linear EW maps presented in Fig.~12(a)-(c) (former numbers) and the 2D~Spec EW maps presented in Fig.~12(d)-(f) (latter numbers). O, Ne, Mg, Si, and Fe abundances are calculated from the abundance maps in Figs~12(g), 13(c), 14(c), 15(c), and 17(b).}\label{Li15bPaperItable:ParaIndividualSpec}
\end{deluxetable}

\section{Erratum: ``XMM-N\lowercase{ewton Large Program on} SN1006 - II: T\lowercase{hermal Emission}" (\href{https://doi.org/10.1093/mnras/stw1640}{2016, MNRAS, 462, 158})} \label{sec:Li16a}

$\bullet$ $n_{\rm e,ISM}$ and $n_{\rm e,ejecta}$ in Table~1 of \citet{Li16a} should be updated to those in Table~\ref{Li16atable:paraexample} of this erratum.

\begin{figure}[!h]
\begin{center}
\epsfig{figure=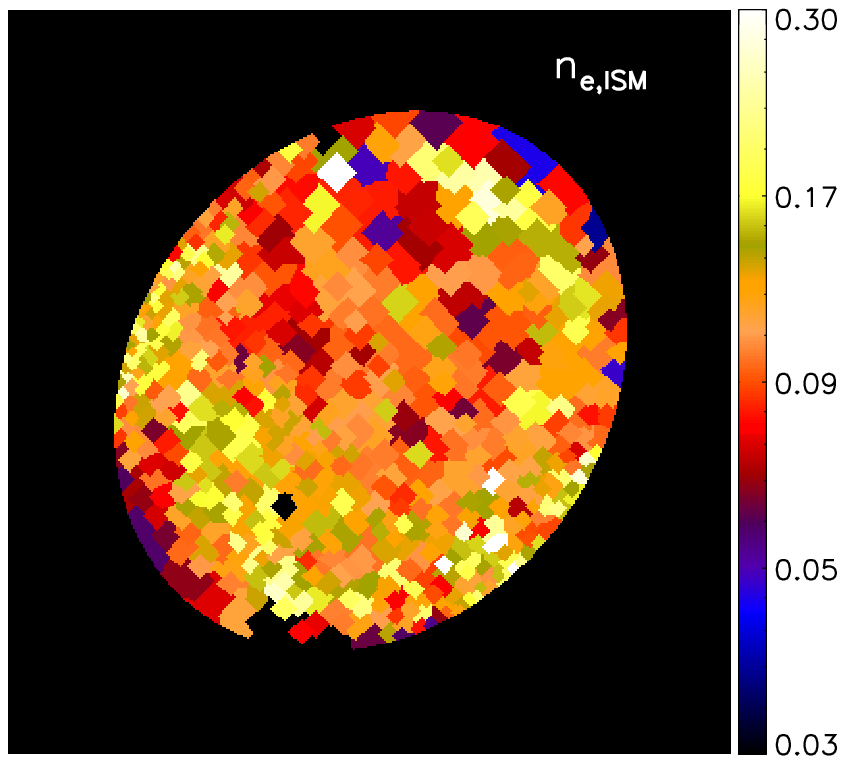,height=0.33\textwidth,angle=0, clip=}
\epsfig{figure=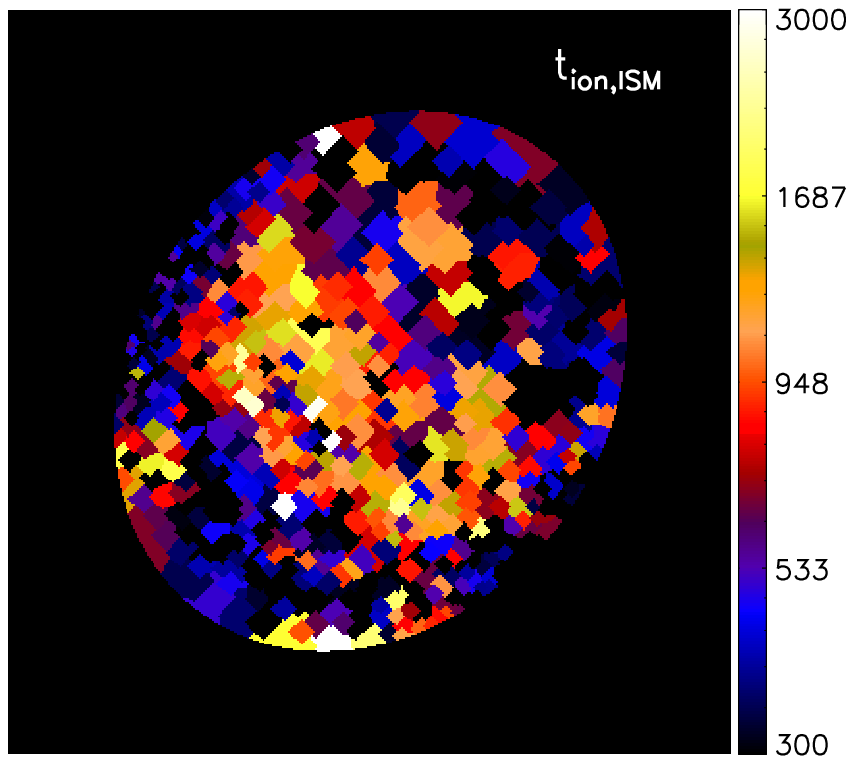,height=0.33\textwidth,angle=0, clip=}
\epsfig{figure=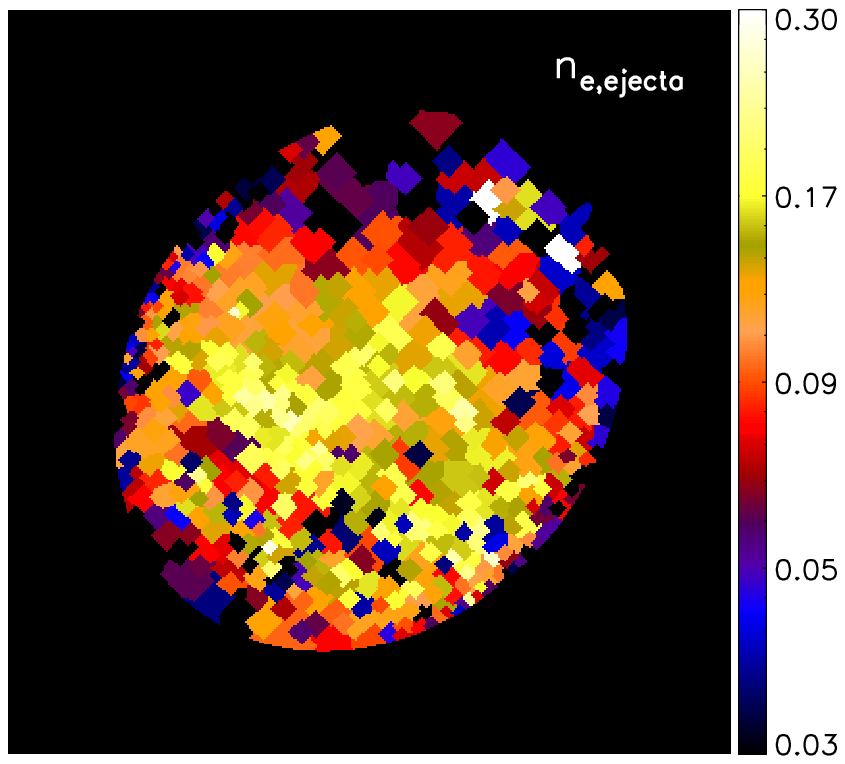,height=0.33\textwidth,angle=0, clip=}
\epsfig{figure=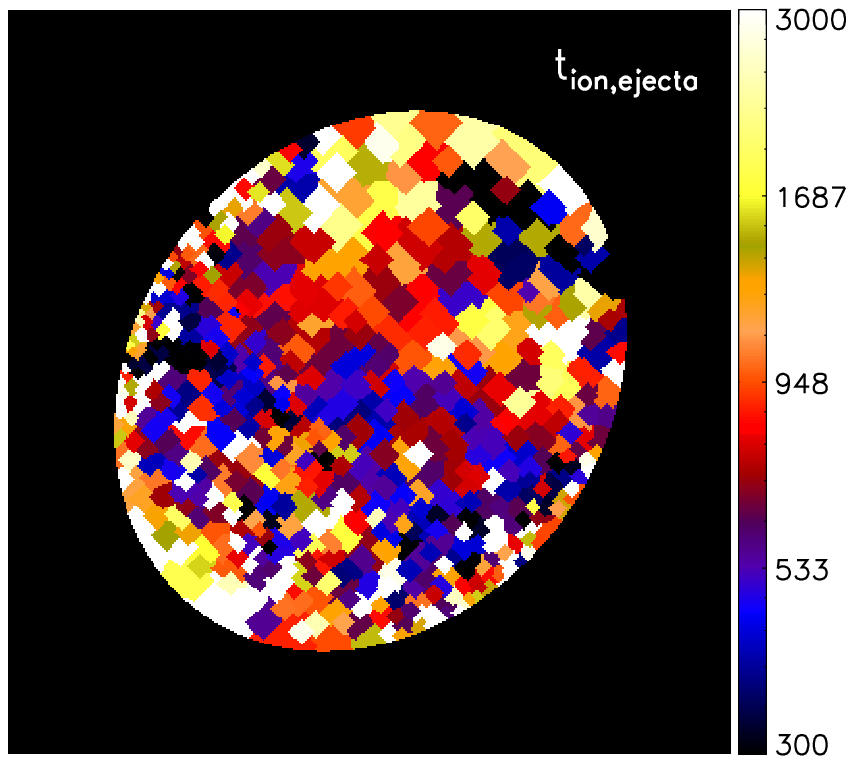,height=0.33\textwidth,angle=0, clip=}
\caption{Updated version of Fig.~4(e), (f), (h), (i) of \citet{Li16a}.}\label{Li16aPaperIIfig:paraimg}
\end{center}
\end{figure}

$\bullet$ Fig.~4(e), (f), (h), (i) of \citet{Li16a} should be updated to the upper left, upper right, lower left, lower right panels of Fig.~\ref{Li16aPaperIIfig:paraimg} of this erratum, respectively. Only the scale of the color bars in these figures are changed.

\begin{table}
\begin{center}
\caption{(Table~1 of \citet{Li16a}) Parameters and errors of the two example regions shown in Fig.~1. Errors are statistical only and are quoted at 90 per cent confidence level.} 
\begin{tabular}{lcccccccccccccc}
\hline
Parameter & reg100515 & reg100573 \\
\hline
$\log (n_{\rm e}t/{\rm cm^{-3}~s})_{\rm ISM}$ & $9.320_{-0.103}^{+0.034}$ & $8.948_{-0.076}^{+0.220}$ \\
$n_{\rm e,ISM}/{\rm cm^{-3}}$ & {\color{red}$0.240_{-0.016}^{+0.022}$} & {\color{red}$0.066_{-0.006}^{+0.009}$} \\
$kT_{\rm ejecta}/{\rm keV}$ & $1.29_{-0.26}^{+0.25}$ & $>9.2$ \\
$\log (n_{\rm e}t/{\rm cm^{-3}~s})_{\rm ejecta}$ & $9.065_{-0.089}^{+0.070}$ & $9.532_{-0.021}^{+0.008}$ \\
$n_{\rm e,ejecta}/{\rm cm^{-3}}$ &  {\color{red}$0.082_{-0.006}^{+0.016}$} & {\color{red}$0.085_{-0.003}^{+0.006}$} \\
$Z_{\rm O,ejecta}/\rm solar$ & $<20$ & $2.39_{-0.33}^{+0.74}$ \\
$(Z_{\rm Ne}/Z_{\rm O})_{\rm ejecta}$ & $0.62_{-0.19}^{+0.21}$ & $0.28_{-0.02}^{+0.03}$ \\
$(Z_{\rm Mg}/Z_{\rm O})_{\rm ejecta}$ & $3.78_{-0.91}^{+0.89}$ & $1.43_{-0.13}^{+0.12}$ \\
$(Z_{\rm Si}/Z_{\rm O})_{\rm ejecta}$ & $13.8_{-2.8}^{+9.7}$ & $3.22\pm0.33$ \\
$(Z_{\rm S}/Z_{\rm O})_{\rm ejecta}$ & $30.1_{-11.6}^{+9.1}$ & $2.21_{-0.95}^{+1.51}$ \\
$(Z_{\rm Fe}/Z_{\rm O})_{\rm ejecta}$ & $<0.39$ & $<0.13$ \\
$v_{\rm ejecta}/\rm (km~s^{-1})$ & $1238_{-344}^{+264}$ & $2818_{-443}^{+75}$ \\
$\alpha$ & $0.11_{-0.01}^{+0.06}$ & $0.1 (<0.103)$ \\
$\nu_{\rm cutoff}/\rm Hz$ & $7.51_{-1.22}^{+0.94}\times10^{14}$ & $6.54_{-0.09}^{+0.78}\times10^{14}$ \\
$\chi^2/\rm d.o.f.$ & 704.14/603 & 1417.55/1258 \\
\hline
\end{tabular}\label{Li16atable:paraexample}
\end{center}
\end{table}

\section{Erratum: ``T\lowercase{he} C\lowercase{ircum}-G\lowercase{alactic} M\lowercase{edium of} MAS\lowercase{sive} S\lowercase{pirals} I: O\lowercase{verview and a Case Study of} NGC 5908" (\href{https://iopscience.iop.org/article/10.3847/0004-637X/830/2/134}{2016, A\lowercase{p}J, 830, 134})} \label{sec:Li16b}

$\bullet$ The following sentences in the abstract of \citet{Li16b} should be changed to:

``Assuming a metallicity of 0.2~solar, an upper limit (without subtracting the very uncertain young stellar contribution) to the mass of hot gas within this radius is {\color{red}$7.3\times10^8\rm~M_\odot$}. The cooling radius is {\color{red}$r_{\rm cool}\approx14\rm~kpc$} or {\color{red}$\approx0.03r_{\rm 200}$}, within which the hot gas could cool radiatively in less than 10~Gyr, and the cooling of hot gas could significantly contribute in replenishing the gas consumed in star formation. The hot gas accounts for {\color{red}$\approx5\%$ of the baryons detected within $r_{\rm 200}$.}'' 

$\bullet$ Table~2 of \citet{Li16b} should be updated to Table~\ref{Li16btable:pararadius} of this erratum.

$\bullet$ The last two paragraphs of Section 4.2 of \citet{Li16b} should be combined together and updated to:

``We further define the cooling radius ($r_{\rm cool}$) as the radius at which $t_{\rm cool}=10\rm~Gyr$, within which the hot gas could radiatively cool and be accreted onto the galaxy. From Equation~(1), we obtain $t_{\rm cool}=10\rm~Gyr$ at $n_{\rm e}\approx6\times10^{-4}\rm~cm^{-3}$. Since the soft X-ray intensity $I_{\rm X}\propto n_{\rm e}^2$, we can estimate the $n_{\rm e}$ distribution from the hot gas soft X-ray intensity profile  (Figure~2), assuming there is no radial variation of the hot gas temperature and metallicity. Adopting the best-fit $\beta$-model of the hot gas component, {\color{red}the estimated $r_{\rm cool}$ and the integrated radiative cooling rate within $r_{\rm cool}$ are $r_{\rm cool}\sim14\rm~kpc$ and $\dot{M}_{\rm hot}\sim0.4~(<1.6)\rm~M_\odot~yr^{-1}$. $\dot{M}_{\rm hot}$, respectively.} Considering the large uncertainties in the estimation of both the hot gas properties and the SFR, we may conclude that the radiative cooling of hot gas could at least significantly contribute in replenishing the gas consumed in star formation.''

$\bullet$ As the baryon budget is only meaningful within a large radial range such as $r_{\rm 200}$ or the virial radius, we change the title of Section 4.3 of \citet{Li16b} to ``{\color{red}Baryon Budget}''. The last paragraph of this section should also be changed to:

``Assuming the filling factor $f=1$, {\color{red}the total baryon mass within the virial radius (described with $r_{\rm 200}$), including the cold atomic and molecular gases, the stars, and the hot gas ($M_{\rm hot}=1.44_{-0.55}^{+3.27}\times10^{10}\rm~M_\odot$), is $M_{\rm b}=2.72_{-0.70}^{+0.77}\times10^{11}\rm~M_\odot$. The hot gas only accounts for a small fraction of the total baryon content ($\approx5\%$). $\sim80\%$ of the expected baryons are not detected as stars or the cold and hot gas.}''

$\bullet$ The following sentences in (2) and (3) of Section~5 of \citet{Li16b} should be changed to:

``(2) In particular, the cooling radius within which the hot gas could cool radiatively within 10~Gyr is {\color{red}$r_{\rm cool}\approx14\rm~kpc$, even smaller than} the outermost radius where the hot gas emission is directly detected. Within this cooling radius, the total radiative cooling rate of the hot gas is {\color{red}$\dot{M}_{\rm hot}\sim0.4~(<1.6)\rm~M_\odot~yr^{-1}$.}''

``(3) Adding the mass of cold atomic and molecular gases, hot gas, and stars, the total baryon mass {\color{red}within the virial radius is $M_{\rm b}\approx2.72_{-0.70}^{+0.77}\times10^{11}\rm~M_\odot$}, dominated by the stellar mass. The hot gas only accounts for {\color{red}$\approx5\%$ of the total baryon content within the virial radius. $\sim80\%$ of the expected baryons are not detected as stars or the cold and hot gas.}''

\begin{table*}
\vspace{-0.in}
\begin{center}
\caption{Inferred parameters of the diffuse hot gas within various radii (Table~2 of \citet{Li16b})} 
\footnotesize
\vspace{-0.0in}
\tabcolsep=4.pt%
\begin{tabular}{lcccccccccccccc}
\hline
            &  $L_{\rm X, 0.5-2keV}$       & $\stretchleftright{\langle}{n_{\rm e}}{\rangle}$  		           & $M_{\rm hot}$                       & $\stretchleftright{\langle}{P_{\rm hot}}{\rangle}$                   & $E_{\rm hot}$                   & $\stretchleftright{\langle}{t_{\rm cool}}{\rangle}$ \\ 
            &  $\rm 10^{39}ergs~s^{-1}$  & $10^{-3}f^{-1/2}\rm cm^{-3}$ & {\color{red}$10^{9}f^{1/2}\rm M_\odot$} & $f^{-1/2}\rm eV~cm^{-3}$ & $10^{57}f^{1/2}\rm ergs$ & $f^{1/2}\rm Gyr$                  \\ 
\hline
$r<15\rm~kpc$ ($1.00^\prime$, $0.036r_{\rm200}$) & $6.83_{-2.20}^{+2.73}$ & {\color{red}$1.95_{-1.38}^{+2.41}$}       & {\color{red}$0.73_{-0.52}^{+0.90}$} & {\color{red}$0.73_{-0.40}^{+2.44}$}       & {\color{red}$0.43_{-0.23}^{+1.43}$} & {\color{red}$3.1~(<6.7)$} \\
$r<25\rm~kpc$ {\color{red}($1.66^\prime$, $0.06r_{\rm200}$)} & $7.09_{-2.28}^{+2.83}$ & {\color{red}$0.93_{-0.66}^{+1.15}$}       & {\color{red}$1.59_{-1.12}^{+1.96}$} & {\color{red}$0.35_{-0.19}^{+1.17}$}       & {\color{red}$0.93_{-0.50}^{+3.11}$} & {\color{red}$6.5~(<14.1)$} \\
$r<50\rm~kpc$ ($3.31^\prime$, $0.12r_{\rm200}$)           & $7.28_{-2.34}^{+2.91}$ & {\color{red}$0.33_{-0.24}^{+0.41}$} & {\color{red}$4.55_{-3.23}^{+5.63}$}           & {\color{red}$0.13_{-0.07}^{+0.42}$} & {\color{red}$2.67_{-1.44}^{+8.90}$}    & {\color{red}$18.0~(<39.3)$} \\
\hline
\end{tabular}\label{Li16btable:pararadius}
\end{center}
\vspace{-0.1in}
The hot gas parameters are scaled based on the best-fit X-ray intensity profile of the hot gas component, without subtracting the young stellar contributions. Errors are 1~$\sigma$ and statistical only. Many systematic uncertainties, such as the poorly constrained metallicity and radial intensity profile at large galactocentric radii, are not included in the errors here. The luminosity $L_{\rm X}$, mass ($M_{\rm hot}$), and thermal energy ($E_{\rm hot}$) are the total values, while the electron number density ($n_{\rm e}$), the thermal pressure ($P_{\rm hot}$), and the radiative cooling timescale ($t_{\rm cool}$) are average values.
\end{table*}

















\acknowledgments JTL and RH would like to acknowledge Prof. Joel Bregman from the University of Michigan and Prof. Daniel Wang from the University of Massachusetts
for helpful discussions. JTL apologizes to the community on the troubles caused by the mistakes discussed in this paper.



\begin{thebibliography}{}
\bibitem[Li et al.(2008)]{Li08} Li J.-T., Li Z. Y., Wang Q. D., Irwin J. A., Rossa J., 2008, MNRAS, 390, 59
\bibitem[Li et al.(2009)]{Li09} Li J.-T., Wang Q. D., Li Z. Y., Chen Y., 2009, ApJ, 706, 693
\bibitem[Li \& Wang(2013)]{Li13a} Li J.-T., Wang Q. D., 2013, MNRAS, 428, 2085
\bibitem[Li et al.(2014)]{Li14} Li J.-T., Wang Q. D., Crain R. A., 2014, MNRAS, 440, 859
\bibitem[Li(2015a)]{Li15a} Li J.-T., 2015a, MNRAS, 453, 1062
\bibitem[Li et al.(2015b)]{Li15b} Li J.-T., Decourchelle A., Miceli M., Vink J., Bocchino F., 2015b, MNRAS, 453, 3953
\bibitem[Li et al.(2016a)]{Li16a} Li J.-T., Decourchelle A., Miceli M., Vink J., Bocchino F., 2016a, MNRAS, 462, 158
\bibitem[Li et al.(2016b)]{Li16b} Li J.-T., Bregman J. N., Wang Q. D., Crain R. A., Anderson M. E., 2016b, ApJ, 830, 134
\bibitem[Li et al.(2017)]{Li17} Li J.-T., Bregman J. N., Wang Q. D., Crain R. A., Anderson M. E., Zhang S., 2017, ApJS, 233, 20
\bibitem[Li et al.(2018)]{Li18} Li J.-T., Bregman J. N., Wang Q. D., Crain R. A., Anderson M. E., 2018, ApJL, 855, 24 
\end{thebibliography}
\end{document}